\documentclass[11pt]{article}
\usepackage{moriond,epsfig}

\def\be{\begin{equation}}
\def\ee{\end{equation}}
\def\bea{\begin{eqnarray}}
\def\eea{\end{eqnarray}}

\newcommand{\bnslash}{\bar n\!\!\!\slash}
\newcommand{\nslash}{n\!\!\!\slash}
\newcommand{\vslash}{v\!\!\!\slash}
\newcommand{\nn}{\nonumber} 
\newcommand{\bn}{{\bar n}}
\newcommand{\bnP}{\bar {\cal P}}
\newcommand{\DSppP}{{{D\!\!\!\!\hspace{0.04cm}\slash}}_c^\perp}
\newcommand{\DSppPu}{{{D\!\!\!\!\hspace{0.04cm}\slash}}_{us}^\perp}

\newcommand{\SCETa}{\mbox{${\rm SCET}_{\rm I}$ }}
\newcommand{\SCETb}{\mbox{${\rm SCET}_{\rm II}$ }}
\newcommand{\mcdot}{\!\cdot\!}

\newcommand{\gtrsim}{\raisebox{-.25em}{$\stackrel{\mbox{\scriptsize $>$}}{\sim}
$}}

\begin{document}

\vspace*{1cm} \title{
THEORETICAL INTRODUCTION TO B DECAYS AND THE SOFT-COLLINEAR
  EFFECTIVE THEORY}

\author{IAIN W. STEWART
    \footnote{\vbox{ \hbox{MIT-CTP-3408.} } 
    Talk given at the XXXVIIIth Rencontres de
    Moriond on QCD and Hadronic Interactions.}
    }

\address{Center for Theoretical Physics, Massachusetts Institute for
Technology,\\ Cambridge, MA 02139 
}

\maketitle\abstracts{ 
  In this talk I give an introduction to methods used to give a model
  independent description of QCD effects in $B$-decays. I discuss three useful
  expansions: $m_b/m_W\ll 1$ for the electroweak Hamiltonian, $\Lambda_{\rm
    QCD}/m_b\ll 1$ for Heavy Quark Effective theory, and $\Lambda_{\rm QCD}/
  Q\ll 1$ for the Soft-Collinear Effective Theory where $Q\sim m_b$ is the large
  energy of light final state hadrons. I discuss predictions that can be made
  with each expansion.
    }

\section{Introduction}

Hadrons containing $b$-quarks are the heaviest known bound states that are
composed entirely of fundamental standard model particles.  The lightest mesons,
$B^0$, $\bar B^0$, $B^{\pm}$, $B_s$ decay weakly, and the plethora of decay
channels provide us with a laboratory for studying electroweak physics and the
hadronic structure of QCD, as well as for searching for physics beyond the
standard model (see Ref.~\cite{reviews} for recent reviews).  These goals go
hand in hand, since from the point of view of studying QCD the electroweak
interactions provide a clean non-interacting probe, while to test the
electroweak interactions or search for new physics the hadronic uncertainties
from QCD must be understood.

Theoretically we wish to simplify as much as possible the role of QCD
interactions without giving up on our ability to make precise predictions. This
is achieved by focusing on the interactions which are strong,
$\alpha_s(\mu\simeq\Lambda)\gtrsim 1$, while expanding in those which are weak
$\alpha_s(\mu\gg\Lambda)\ll 1$. This requires distinguishing the perturbative
mass scales $m_W, m_b, \ldots$, from $\Lambda_{\rm QCD}$, and is made possible
by constructing effective field theories.  The ratio of scales serve as small
expansion parameters so that predictions can be improved to any desired
precision, albeit at the cost of additional non-perturbative hadronic
parameters.  In this talk I focus on describing the QCD aspects of $B$-decays
inputing bits of electroweak physics as necessary.  In particular I discuss the
separation of the scales $m_W,m_t$ (mediating the weak decays), $m_b,Q$, (the
mass of the decaying particle and energy released to light final state
particles), and $\Lambda_{\rm QCD}$ (the scale responsible for binding in the
hadrons).

\section{The Electroweak Hamiltonian, $m_b\ll m_W$}

By integrating out the top quarks and $Z$ and $W$ bosons from the standard model
we arrive at an effective weak Hamiltonian $H_W$ describing interactions
involving the $b$ and lighter quarks.  Since the expansion parameter is very
small $m_b/m_W \sim 1/17$ it suffices to work to leading order (meaning first
order for semileptonic decays, second order for $B$--$\bar B$ mixing etc.). In particular for non-leptonic $\Delta
B=\pm 1$ transitions we have
\begin{eqnarray} \label{Hw}
  H_W^{\Delta B=\pm 1} &=&  \frac{G_F}{\sqrt{2}} \sum_{j,k,\ell=1}^2 \sum_{i} 
  V_{u_j b}^{\phantom{*}} V_{u_k d_\ell}^*
   C_i(\mu)\: O_i(\mu) + {\rm h.c.} \,.
\end{eqnarray}
The operator basis can be reduced using the equations of motion to
\begin{eqnarray} \label{Oi}
  O_1 &=& (\bar u_j b)_{V-A} (\bar d_\ell u_k)_{V-A} \,, \qquad\ \quad\
  O_2 = (\bar u_j^\alpha b^\beta)_{V-A} (\bar d_\ell^\beta u_k^\alpha)_{V-A}  
    \,, \nn\\
  O_3 &=& \delta_{jk}\, (\bar d_\ell b)_{V-A} \sum_{q} (\bar q q)_{V-A} \,,
  \qquad
  O_4 = \delta_{jk}\, (\bar d_\ell^\alpha b^\beta)_{V-A}
     \sum_{q} (\bar q^\beta  q^\alpha)_{V-A}
     \,, \nn\\
  O_5 &=& \delta_{jk}\, (\bar d_\ell b)_{V-A} \sum_{q} (\bar q q)_{V+A} \,,
  \qquad
  O_6 = \delta_{jk}\, (\bar d_\ell^\alpha b^\beta)_{V-A} 
     \sum_{q} (\bar q^\beta  q^\alpha)_{V+A}
     \,, \nn\\
  O_7 &=& -\frac{e\, m_b}{16\pi^2}\:\delta_{jk}\,
    \bar d_\ell \sigma^{\mu\nu} F_{\mu\nu} P_R\, b 
    \,, \qquad
  O_8 = -\frac{g\, m_b}{16\pi^2}\:\delta_{jk}\, 
     \bar d_\ell \sigma^{\mu\nu} G_{\mu\nu} P_R\, b 
     \,,
\end{eqnarray}
with flavors $d_1=d$, $d_2=s$, $u_1=u$, $u_2=c$, $q=\{u,d,s,c,b\}$, and color
indices $\alpha,\beta$. Here $O_{1,2}$ are current-current operators,
$O_{3,4,5,6}$ are QCD penguin operators, and $O_{7,8}$ are magnetic penguins.
Additional electroweak penguin operators $O_{7,8,9,10}^{ew}$ are also relevant
but are not shown.  For the $\Delta B=\pm 2$ operators relevant to $B$--$\bar B$
mixing we have
\begin{eqnarray}
  H_W^{\Delta B=\pm 2} &=&  \frac{G_F^2}{16\pi^2} \sum_{\ell=1}^2 
  (V_{t b}^{\phantom{*}} V_{t d_\ell}^*)^2
   C_{B\bar B}(\mu)\: [(\bar d_\ell b)_{V-A} (\bar d_\ell b)_{V-A}](\mu)
   + {\rm h.c.}
\end{eqnarray}
Integrating out the short distance physics associated with the scales $m_W,m_t$
has left us with products of CKM parameters $V_{u_j b}^{\phantom{*}} V_{u_k
  d_\ell}^*$, along with Wilson coefficients $C_i(\mu)=C_i(\mu,m_Z,m_t)$ which
encode QCD interactions that take place at scales $\mu> m_b$.  The operators
$O_i$ encode the remaining long distance physics associated with the scales
$\mu=m_b, m_c, m_{s,d,u}, \Lambda_{\rm QCD}$.  The computation of the $C_i$ can
be carried out with free external quarks and light quark fields expanded about
their massless limit. This follows from the fact that matching calculations are
independent of our choice of states and the $C_i$ do not depend on the long
distance scales.  To evolve the operators from the short distance scale
$\mu=m_W$ down to $\mu=m_b$ we make use of the renormalization group which
simultaneously sums series of large logarithms $[\alpha_s \ln(m_W^2/m_b^2)]\sim
1$ into the coefficients $C_i(\mu)$. Finally, the matrix elements of $O_i$
between physical hadronic states still fully depend on the long distance scales,
and the quarks and gluon fields in these operators interact through the full QCD
Lagrangian.


The result for $H_W$ gives some useful predictions even without distinguishing
the long distance scales any further. One example is in $\bar B^0$-$B^0$ 
($\bar B_s$-$B_s$) mixing for $\Delta m_d$ ($\Delta m_s$), the difference
between mass eigenstates.  The box diagrams are dominated by top quarks, and
\begin{eqnarray}
  \Delta m_q = \frac{G_F^2 m_W^2 m_{B_q}}{6\pi^2}\bigg\{
  |V_{tb}^{\phantom{*}} V_{tq}^*|^2 \Big[  S(x_t) \eta_B f(\mu) \Big]\ 
  f_{B_q}^2 B_{B_q}(\mu) + {\cal O}\Big(\frac{m_b^2}{m_W^2}\Big) \bigg\}
\end{eqnarray}
where the Inami-Lin function $S(x_t)$ contains the dependence on
$x_t=m_Z^2/m_t^2$ from integrating out the top and electroweak bosons. The
product $\eta_B f(\mu)$ contains computable QCD effects from scales between
$m_W$ and $m_b$.  Finally, the product $f_{B_q}^2 B_{B_q}$ arises from the long
distance matrix element $\langle B^0 | O^{\Delta B=\pm 2} | \bar B^0 \rangle$.
These matrix elements can be computed using lattice QCD without worrying about
also encoding physics at the $W$-scale.

\begin{figure}[t!]
\vskip0.1cm
 \centerline{
  \mbox{\epsfxsize=6truecm \hbox{\epsfbox{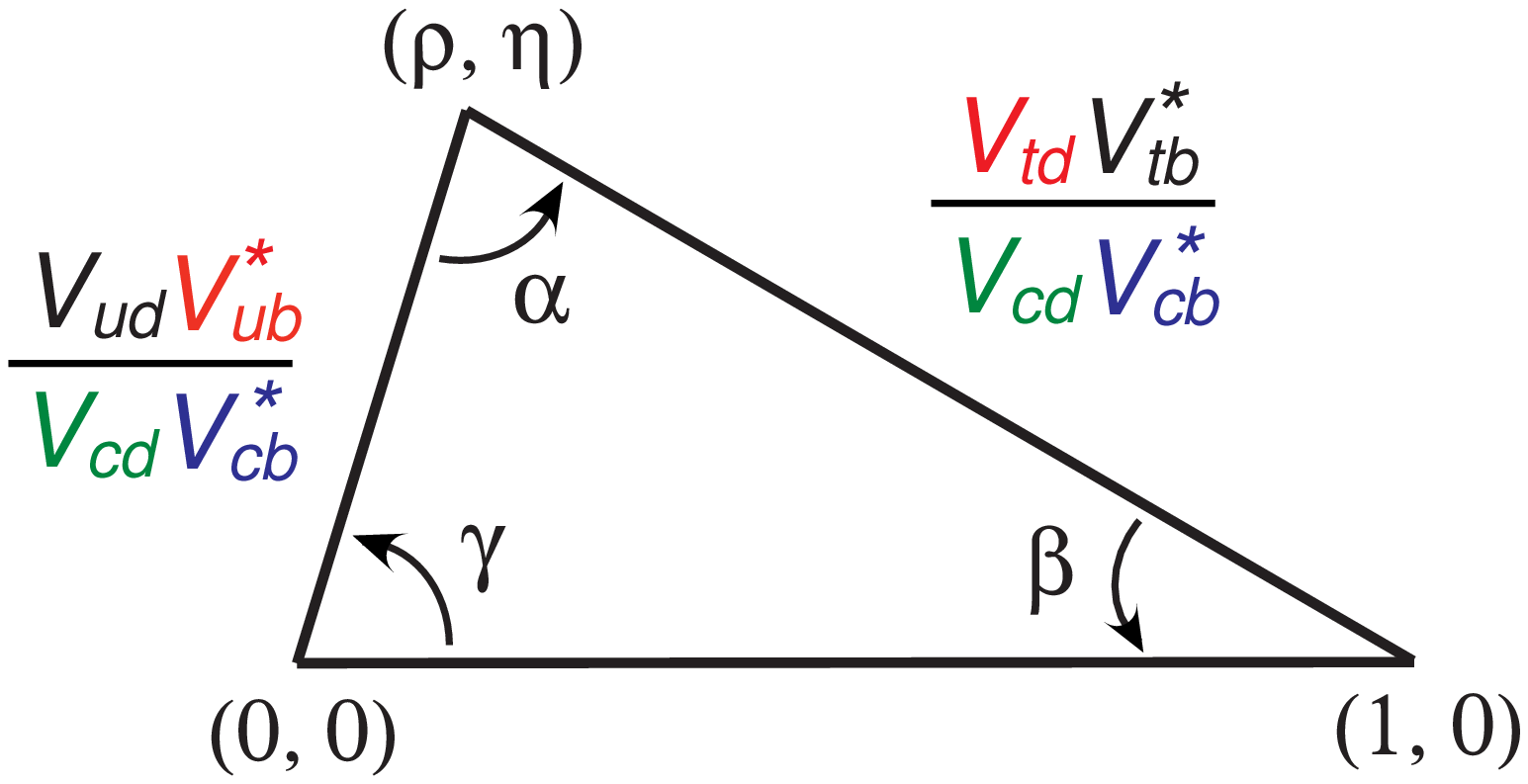}} }
  \mbox{\epsfxsize=8truecm \hbox{\epsfbox{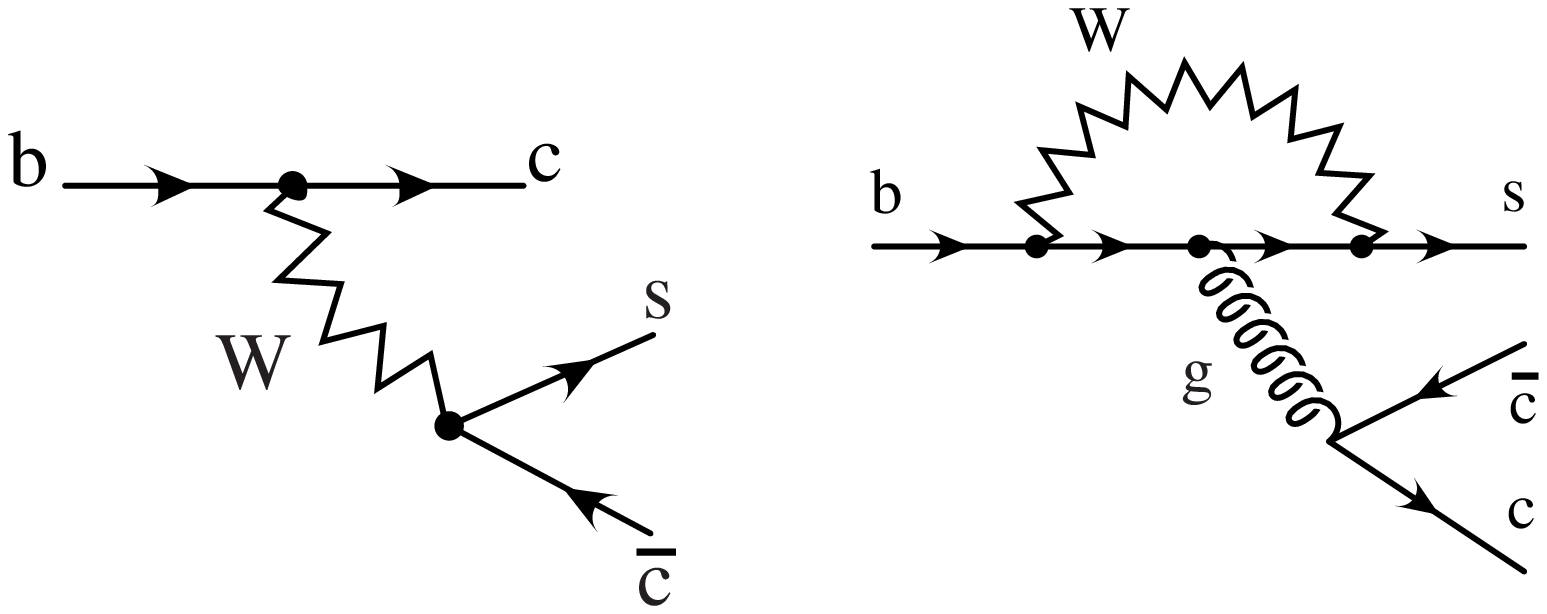}} }
   } 
  \caption{Unitarity triangle, and examples of standard model tree and penguin 
  diagrams.}\label{fig_ut}
\end{figure}
As our second example we consider the $\sin(2\beta)$ measurement of CP-violation
using time dependent CP asymmetries like~\cite{Bigi}
\begin{eqnarray}
  a_{CP}(t) = \frac{\Gamma[\bar B^0(t) \to J/\psi K_S] 
   - \Gamma[ B^0(t) \to J/\psi
    K_S]}{\Gamma[\bar B^0(t) \to J/\psi K_S] 
   + \Gamma[ B^0(t) \to J/\psi K_S]} \,.
\end{eqnarray}
In the standard model CP-violation occurs through the CKM matrix, and is often
envisioned in the unitary triangle shown in Fig.~\ref{fig_ut}.  To see why the
measurement of $\sin(2\beta)$ is theoretically clean, note that the amplitude
for $\bar B^0(t) \to J/\psi K_S$ has the form
\begin{eqnarray} \label{Abar}
 \bar A(\bar B^0(t) \to J/\psi K_S) = (V_{cb}^{\phantom{*}}
   V_{cs}^{*}+V_{ub}^{\phantom{*}} V_{us}^*)
   \: P_t + V_{cb}^{\phantom{*}} V_{cs}^* (P_c +T)
   + V_{ub}^{\phantom{*}} V_{us}^* P_u \,.
\end{eqnarray}
In running below $m_W$ the operators $O_{1,2}$ mix with the penguin operators
$O_{3-6}$, but do not upset the CKM structure.  At a scale $\mu\simeq m_b$ the
matrix elements in Eq.~(\ref{Abar}) are
\begin{eqnarray}
 && P_c + T = \sum_{i=1,2}C_{i}\, \big\langle  O_{i}(j,k,\ell=2,2,2)\big\rangle 
   \,,\qquad
 P_t = - \sum_{i=3,\ldots} C_{i} \, \big\langle O_{i}(\ell=2,\mbox{ any q})
  \big\rangle \,,\nn\\
 && \qquad P_u = \sum_{i=1,2} C_{i} \, \big\langle O_{i}(j,k,\ell=1,1,2)
  \big\rangle \,.
\end{eqnarray}
The ``T'' and ``P'' notation denotes the fact that we often picture the $b\to
c\bar c s$ transition that takes place in the matrix elements as occurring
through ``tree'' or ``penguin'' diagrams as shown in
Fig.~\ref{fig_ut}.~\footnote{ This notation can be confusing. For instance the
  $O_1(j,k,\ell)=O_1(2,2,2)$ operator is generated by a tree topology at lowest
  order at $m_W$ but has both tree and penguin contractions in the long distance
  matrix element. On the other hand $O_1(1,1,2)$ is generated by a tree topology
  but has only penguin contractions for $B\to J/\psi K_S$.}  Since
$|V_{ub}^{\phantom{*}}V_{us}^*|/|V_{cb}^{\phantom{*}}V_{cs}^*|\sim 1/50$ the
amplitude $\bar A$ is dominated by a single combination of strong matrix
elements, $P_t+P_c+T$.  In the ratio of $\bar A$ to its CP-conjugate amplitude
$A$ any strong phase generated in $P_t+P_c+T$ cancels exactly. This uses the
fact that QCD is known to preserve CP at the level of one part in $10^{12}$.
Thus, the asymmetry $a_{CP}(t) = \sin(2\beta) \sin(\Delta m_d\: t)$ is dominated
by the weak phase $\beta$.  

\section{The Heavy Quark Effective Theory, $\Lambda_{\rm QCD}\ll m_b$}

To improve our description of matrix elements we can consider separating the
scale $m_b$ (and $m_c$) from the lighter scales $m_{u,d,s}$ and $\Lambda_{\rm
  QCD}$ using the heavy quark effective theory (HQET).~\cite{hbook} This is
accomplished by integrating out the heavy anti-quarks and only keeping heavy
quarks with fluctuations close to their mass-shell, $p_b=m_b v+k$ with $k\sim
\Lambda_{\rm QCD}$ and $v^2=1$. The QCD heavy quark fields $Q=b$ ($Q=c$) are
traded for HQET fields, $Q(x) = \sum_v e^{-i m_Q v\cdot x} h_v^{(Q)}(x)$ with
$\vslash h_v^{(Q)} =h_v^{(Q)}$, and all operators involving a $b$ ($c$) develop
an expansion in $1/m_Q$.  The Lagrangian describing the heavy quarks is also
expanded,
\begin{eqnarray} \label{LHQET}
  {\cal L}_{\rm HQET} &=& \sum_{v,Q}\ \bar h_v^{(Q)} iv\mcdot D h_v^{(Q)}
   +  \frac{1}{2m_Q} \bar h_v^{(Q)} \Big[ (i D_\perp)^2 
   + \frac{ c_F(\mu)}{2} g\,\sigma_{\mu\nu}  G^{\mu\nu} \Big] h_v^{(Q)} 
   + \ldots \,.
\end{eqnarray}
In HQET Wilson coefficients like $c_F(\mu)$ now incorporate QCD effects from
scales $m_Q\, \gtrsim\, \mu \gg \Lambda_{\rm QCD}$. At lowest order ${\cal
  L}_{\rm HQET}$ has an $SU(4)$ spin-flavor symmetry not present in QCD which
leads to additional predictions (such as relating the states $B,B^*,D$, and
$D^*$).  Constructing HQET allows us to separate perturbative
$\alpha_s(m_b)$ effects and define universal hadronic parameters.  An example is
$\bar\Lambda$, $\lambda_{1}$, $\lambda_2$ which appear in the heavy meson mass
formula $m_{B^{(*)}} = m_b +\bar\Lambda -\lambda_1/(2m_b) + d_h c_F(\mu)
\lambda_2(\mu)/(2 m_b)+\ldots$, and are defined by matrix elements of terms in
the series in Eq.~(\ref{LHQET}).

The HQET expansion is very useful in making predictions.  In fact we have
already encountered one example, for $f_{B_q}^2 B_{B_q}$ lattice calculations
are more practical using HQET for $b$ quark fields (or the related
Non-Relativistic QCD effective theory).  This makes it possible to use larger
lattice spacings $a\gg 1/m_b$ and obtain more statistics to probe the
non-perturbative effects at distance scales $\sim 1/\Lambda_{\rm QCD}$.  Another
well known application of HQET is measuring $|V_{cb}|$ with $B\to D^*
\ell\bar\nu_\ell$. The exclusive differential decay rate is
\begin{eqnarray} \label{excl}
  \frac{d\Gamma(B\to D^*\ell\bar\nu_\ell)}{d\omega}
  &=& \frac{G_F^2|V_{cb}|^2 m_B^5}{48\pi^3} 
    \sqrt{\omega^2-1}\: f(r^*,\omega)\ {\cal F}_{*}^2(\omega) 
\end{eqnarray}
where $r_*=m_{D^*}/m_B$, phase space produces $f(r^*,\omega)=r_*^3 (1-r_*)^2
(\omega+1)^2 [ 1 + 4\omega(1-2\omega r_* + r_*^2)/(1+\omega)/(1-r_*)^2 ]$, and
${\cal F}_*(\omega)$ contains the dependence on four QCD form factors.  At
leading order in $\Lambda_{\rm QCD}/m_{b,c}$ this dependence collapses to just
the Isgur-Wise function, ${\cal F}_*(\omega)=\xi(\omega)$ which is normalized at
zero recoil due to heavy quark symmetry, $\xi(1)=1$.  Experimentally
extrapolating Eq.~(\ref{excl}) to zero recoil therefore gives us a method of
measuring $|V_{cb}|$. Corrections to ${\cal F}_*(1)=1$ include calculable
$\alpha_s(m_b)$ terms in Wilson coefficients, power corrections of order
$1/m_Q^2$ which are amenable to determination on the lattice (order $1/m_Q$
effects vanishing by Luke's theorem), and effects of the shape of ${\cal
  F}_*(\omega)$ which are constrained by analyticity.~\cite{pert}  HQET can also
be used for inclusive decays such as $B\to X_c \ell\bar\nu_\ell$ and $B\to
X_u\ell\bar\nu_\ell$.\cite{inclrev} The decay rates can be computed using an
operator product expansion and beyond leading order depend on the matrix elements
of HQET operators, $\bar\Lambda$, $\lambda_1$, and $\lambda_2$.  Together with
$|V_{cb}|$ these parameters must be simultaneously fit to the data (the fit to
$m_{B^*}-m_B$ gives $\lambda_2=0.128\pm 0.010\,{\rm GeV}^2$). Taking the average
of the CLEO and BaBar results\cite{Vcbexpt0} I find
\begin{eqnarray}
  \bar\Lambda^{\overline {\rm MS}} = 440 \pm 80 \: {\rm MeV}\,,\qquad
  \lambda_1^{\overline {\rm MS}} = -0.299 \pm 0.069 \: {\rm GeV}^2 
   = - ( 547 \pm 64 \,{\rm MeV} )^2 \,.
\end{eqnarray}
The current averages for $|V_{cb}|$ from the heavy flavor averaging group
~\cite{Vcbexpt1} are
\begin{eqnarray}
 |V_{cb}|^{\rm excl} = ( 42.6 \pm 1.2 \pm 1.9) \times 10^{-3} \,,\qquad
 |V_{cb}|^{\rm incl} = ( 41.9 \pm 0.7 \pm 0.6) \times 10^{-3} \,,
\end{eqnarray} 
where the first errors are experimental and the second are an estimate of the
remaining theoretical uncertainty.

A heavy quark expansion is also useful in determinations of $|V_{ub}|$ from
semileptonic $b\to u$ decays. For these decays the cuts to exclude the large
$b\to c$ background make things more challenging. For $B\to X_u\ell\bar\nu_\ell$
there are four regions where different theoretical tools apply:
\begin{eqnarray}
\begin{tabular}{lllll}
 I) & &  $m_X^2 \gg E_X \Lambda \gg \Lambda^2$ & & local OPE in $\Lambda/m_b$
   , \\
 II) & &  $m_X^2 \sim E_X \Lambda \gg \Lambda^2$ & & shape function region 
   , \\
 III) & &  $m_X^2 \sim \Lambda^2$, $E_X\sim \Lambda$ & & resonance region 
   ($B\to \pi\ell\bar\nu_\ell$, $\ldots$), slow exclusive $X$, \\
 IV) & &  $m_X^2 \sim \Lambda^2$, $E_X\gg \Lambda$ & & resonance region,
   energetic exclusive $X$. \nn
\end{tabular} 
\end{eqnarray}
Region I is the most theoretically clean, and involves a local OPE as in $B\to
X_c \ell\bar\nu_\ell$ decays. Cuts on $m_X^2$ and $q^2$ of the leptons can be
made to reject the $b\to c$ background and remain in this region.\cite{cuts}
Another possibility is to experimentally reconstruct the total rate.  The
simplest cut which rejects $b\to c$ is on the charged lepton energy and puts us
in region II.  Typically cuts which reject $b\to c$ and leave us in II have more
events. In this region the rate becomes sensitive to a non-perturbative shape
function and has an expansion in $\Lambda/E_X$. The shape function can be
measured in $B\to X_s\gamma$, which then allows for a model independent
extraction of $|V_{ub}|$ in the endpoint region.  For regions III and IV we are
dealing with exclusive decays and a heavy quark expansion alone does not provide
enough information for a model independent extraction of $|V_{ub}|$.  The
theoretical description of regions II and IV can be improved by using the
soft-collinear effective theory which is discussed in the next section.
Currently for exclusive decays the experimental analysis uses form factor models
and QCD sum rules~\cite{Ball} and the theoretical uncertainty is still quite
large (but should improve as lattice results become more accurate).  Averaging
the BaBar, Belle, and CLEO results~\cite{Vubexpt,Vcbexpt1} separately for each
region avoids dealing with their quantatively different uncertainties and gives
\begin{eqnarray}
 && |V_{ub}|^{\rm II}_{\rm M_X^2} = ( 4.21 \pm 0.29\pm 0.55 ) \times 10^{-3} \,,
  \qquad
 |V_{ub}|^{\rm II}_{\rm endpt} = ( 4.17 \pm 0.14 \pm 0.6 )\times 10^{-3} \,, \nn\\
 && 
 |V_{ub}|^{\rm III + IV} = ( 3.4 \pm 0.2 \pm 0.5) \times 10^{-3} \,.
\end{eqnarray} 
The uncertainties are experimental and an estimate of the theoretical
uncertainty (the latter values are taken from those quoted by
the experimental analyses with no reduction for averaging).

\section{The Soft-Collinear Effective Theory, $\Lambda_{\rm QCD}\ll Q$}

To improve the theoretical description of $B$-decay processes with energetic
hadrons we can consider separating the scales $Q\simeq E_X, m_b$ from the
lighter scales $m_{u,d,s,}$, $\Lambda_{\rm QCD}$ using the soft-collinear
effective theory (SCET).\cite{bfl,bfps,cbis,bps2} This theory includes all the
degrees of freedom of HQET, but adds collinear quarks and gluons with large
energy $p^-\sim Q$, but small offshellness $p^2\sim Q^2\lambda^2$ where
$\lambda\ll 1$. SCET describes decays or scattering processes with energetic
hadrons or jets, and can be used to describe $B$-decays to light hadrons as well
as more traditional QCD processes like DIS or form factors.  The analog of
the scale separation formula in Eq.~(1) are now more complicated and are
referred to as factorization theorems. They involve convolutions between
perturbative and non-perturbative functions.  The convolutions arise because the
separation of scales $Q^2\gg Q\Lambda\gg \Lambda^2$ allows perturbative and
non-perturbative momenta $p^2$ to communicate in some components. For example
$p^-$ collinear momenta with hard $p^2\sim Q^2$ momenta.  The goal of model
independent analysis in SCET is the same as that in HQET, namely to measure the
non-perturbative functions in one process and then use them in another.

In the remainder I give a brief overview of the SCET formalism focusing on two
sets of degrees of freedom, \SCETa (collinear $p^2\sim Q\Lambda$, usoft $p^2\sim
\Lambda^2$) and \SCETb (collinear $p^2\sim \Lambda^2$, soft $p^2\sim\Lambda^2$).
The former theory describes processes with inclusive jets of size $p_\perp^2\sim
Q\Lambda$ and soft hadrons, while the latter describes exclusive processes with
both energetic and soft hadrons.  Using light cone coordinates ($p^+=n\cdot p$,
$p^-=\bn\cdot p$, $p^\perp$), the momenta of collinear fields scale as
$p^\mu\sim Q(\lambda^2,1,\lambda)$, $\lambda=\sqrt{\Lambda/Q}$ in \SCETa, while
$p^\mu\sim Q(\eta^2,1,\eta)$ with $\eta=\Lambda/Q$ in \SCETb.

In \SCETa an important ingredient is the lowest order collinear quark Lagrangian
\begin{eqnarray}\label{Lc0}
{\cal L}_{\xi\xi}^{(0)} = \bar \xi_{n,p'} \left\{in\cdot D \!+\!
 i\DSppP \frac{1}{i\bn\!\cdot D_c} i\DSppP \right\} 
 \frac{\bnslash}{2}\: \xi_{n,p} \,,
\end{eqnarray}
where $in\mcdot D = i n\cdot \partial + g n\cdot A_n + g n\cdot A_{us}$,
$i\bn\cdot D_c=\bnP+g\bn\cdot A_{n}$, $ iD_c^{\perp\mu} = \bnP_\perp^\mu +
gA_{n}^{\perp\mu}$. It has interactions with all components of the collinear
gluons and includes collinear quark pair production and annihilation.  For
interactions with usoft gluons $A_{us}$ only the $n\cdot A_{us}$ components
interact with quarks at leading order and the collinear propagators reduce to
eikonal propagators when the external collinear quarks are put on-shell.

In deriving the free propagator from Eq.~(\ref{Lc0}) it naively appears that
only a single pole would appear in the complex plane, $1/(n\mcdot p +
p_\perp^2/\bn\mcdot p + i\epsilon)$, which would indicate that ${\cal
  L}_{\xi\xi}^{(0)}$ does not contain both particles and antiparticles. To
arrive at the correct result~\cite{cbis} we recall that the field $\xi_{n,p}=
\xi_{n,p}^+ + \xi_{n,-p}^-$ so a negative $\bn\mcdot p$ momentum label
corresponds to an antiparticle with $-\bn\mcdot p>0$.  Taking label momentum
$(\bn\mcdot p,p_\perp)$ and residual momentum $n\mcdot p$ both in the direction
of the fermion arrow, the full SCET propagator is
\begin{eqnarray}
  i\frac{\nslash}{2} \bigg[
  \frac{\theta(\bn\mcdot p)}{n\mcdot k + p_\perp^2/\bn\mcdot p + i\epsilon }
 -\frac{\theta(-\bn\mcdot p)}{-n\mcdot p - p_\perp^2/\bn\mcdot p + i\epsilon }
  \bigg] \,.
\end{eqnarray}
The second term corresponds to the collinear antiquarks. In a more compact 
form we have
\begin{eqnarray} \label{cqprop}
 i\frac{\nslash}{2} \bigg[
 \frac{\theta(\bn\mcdot p)}{n\mcdot p + p_\perp^2/\bn\mcdot p + i\epsilon }
 +\frac{\theta(-\bn\mcdot p)}{n\mcdot p + p_\perp^2/\bn\mcdot p - i\epsilon }
 \bigg]
= i\frac{\nslash}{2} \: \frac{\bn\mcdot p }
   {n\mcdot p\: \bn\mcdot p + p_\perp^2 + i\epsilon } \,.
\end{eqnarray}
Thus there are indeed two poles in the collinear quark propagator.  An
alternative way of writing the propagator in Eq.~(\ref{cqprop}) is
\begin{eqnarray}
  \frac{1 }
   {n\mcdot p\:  + p_\perp^2/\bn\mcdot p + i\epsilon\: {\rm sign}(\bn\mcdot p) } \,.
\end{eqnarray}
This form is useful for considering interactions with ultrasoft gluons. If a
collinear quark propagator has momentum $p+k$ where $k$ is ultrasoft and $p$
external, then onshell $n\mcdot p + p_\perp^2/\bn\mcdot p=0$ and the propagator
reduces to 
\begin{eqnarray}
  \frac{1 }
   {n\mcdot k\:  + i\epsilon\: {\rm sign}(\bn\mcdot p) } \,,
\end{eqnarray}
which is eikonal as promised. If SCET is formulated in position
space~\cite{bcdf,HN}, then separate fields for quarks and antiquarks can be used
to construct ${\cal L}_{\xi\xi}^{(0)}$ and obtain this pole structure.

The form of operators and interactions in SCET are constrained by symmetries.
For the spin structure note that $\xi_n$ are two-component fields with $\nslash
\xi_n=0$. The spinor components which are small for motion in the $n$ direction
have been integrated out. The ${\cal L}_{\xi\xi}^{(0)}$ Lagrangian has a global
helicity spin symmetry~\cite{bfps} with generator $h_n=\epsilon_\perp^{\mu\nu}
[\gamma_\mu,\gamma_\nu]/4$. $h_n$ is independent of individual collinear
particle momenta since $\xi_{n,p}$ only describe particles moving close to the
$n$ direction.  SCET also has a rich set of gauge symmetries.\cite{bps2} In
particular both the collinear and usoft gluons have their own set of gauge
transformations which have support over the corresponding momenta,
$i\partial^\mu U_c\sim (\lambda^2,1,\lambda)$ and $i\partial^\mu U_u \sim
(\lambda^2,\lambda^2,\lambda^2) $. The longer distance $A_{us}^\mu$ fields act
like background fields to the shorter distance collinear gluon $A_n^\mu$ fields.
Lorentz symmetry is also realized in an interesting way in SCET. The
introduction of the basis vectors $n$ and $\bn$ breaks some of the symmetry,
which is then restored by reparameterization transformations which leave
$n\mcdot \bn=2$ and $n^2=\bn^2=0$.\cite{rpi} These transformations fall into
three classes, I) $n\to n+\Delta_\perp$, $\bn\to\bn$, II) $n\to n$, $\bn\to
\bn+\epsilon_\perp$, and III) $n\to e^\alpha n$, $\bn\to e^{-\alpha}\bn $, where
$\Delta_\perp\sim \lambda$ and $\epsilon_\perp,\alpha\sim \lambda^0$. Demanding
invariance under these transformations order by order in $\lambda$ eliminates
some collinear operators and constrains the form of others, see for
example.\cite{bcdf,ps1} A final constraint on operators which applies only to
\SCETa is locality. \SCETa has non-local objects such as the Wilson line $W$
built out of $\bn\mcdot A_n$ fields and $\bn\mcdot p$ momenta which are $\sim
Q$, however no inverse powers of smaller momenta appear in Feynman rules.  For
\SCETb we integrate out the scale $Q\Lambda$, and operators involving soft and
collinear fields are more non-local as emphasized in Ref.~\cite{HN}.

An important aspect of SCET is the factorization of distance scales.
Hard-collinear factorization is codified in the relation $C(i\bn\mcdot D_c)=W
C(\bnP) W^\dagger$, where $\bnP$ is a momentum operator that picks out
$\bn\mcdot p$ momenta of collinear fields.\cite{cbis} For example suppressing
power corrections it allows us to write the QCD heavy-to-light current 
\begin{eqnarray}
  \bar u\Gamma b &=& \bar\xi_n C(-i\bn\mcdot \overleftarrow D_c) \Gamma (W h_v)
  = [\bar\xi_n W \Gamma h_v]\: C(\bnP^\dagger)  \nn\\
 &=& \int d\omega\: (\bar\xi_n W)_\omega\, \delta(\omega-\bnP^\dagger) \Gamma h_v 
  \, C(\omega)
  = \int d\omega\: J(\omega) \: C(\omega)\,,
\end{eqnarray}
which separates the perturbatively calculable coefficient $C(\omega)$ from the
longer distance SCET current $J(\omega)$. Another type of factorization,
usoft-collinear, occurs through interactions within SCET. By making the field
redefinitions~\cite{bps2} $\xi_n\to Y\xi_n$, $A_n^\mu\to Y A_n^\mu Y^\dagger$
the coupling of usoft gluons is removed from the leading quark and gluon
Lagrangians ${\cal L}_{\xi\xi}^{(0)}$, ${\cal L}_{cg}^{(0)}$. Here $Y$ is a path
ordered exponential along the $n$ direction from $-\infty$ to $x^\mu$ of
$n\mcdot A_{us}$ gluons. After the field redefinition, factors of $Y$ appear in
operators involving usoft fields such as the heavy-to-light current $(\bar\xi_n
W)_\omega \Gamma Y^\dagger h_v$. They also show up in power suppressed operators
such as the order $\lambda$ collinear quark Lagrangian
\begin{eqnarray}
 {\cal L}_{\xi\xi}^{(1)} &=& (\bar\xi_n W) (Y^\dagger i\DSppPu Y) \frac{1}{\bnP}
   (W^\dagger i\DSppP W) (W^\dagger \xi_n) + \mbox{h.c.}
\end{eqnarray}
Using relations such as $(Y^\dagger i \DSppPu Y) = i\partial_{us}^\perp +
[1/(in\mcdot\partial) Y^\dagger [in\mcdot D_{us}, i \DSppPu] Y]$ it always
possible to collect the collinear and usoft fields in separate matrix elements
even at subleading order in $\lambda$.

To construct \SCETb it is useful to make use of results for \SCETa, essentially
because \SCETa describes how the soft-collinear modes of \SCETb
communicate.~\cite{bps4} After factorizing the usoft-collinear interactions in
\SCETa we match \SCETa to \SCETb by integrating out $p^2\sim Q\Lambda$
fluctuations. This is done using states involving only $p^2\sim \Lambda^2$
particles. In many cases this matching is trivial, for instance ${\cal
  L}_{\xi\xi}^{(0)}\to {\cal L}_{\xi\xi}^{(0)}$ and $\bar\xi_n W\Gamma Y^\dagger
h_v\to \bar\xi_n W \Gamma S^\dagger h_v$ (where $S$ is an identical Wilson line
to $Y$ but built out of soft fields). In external operators and at subleading
order in the power counting matching calculations involving time ordered
products of operators with both collinear and usoft fields appear. These can
generate additional Wilson coefficients, and are referred to as jet functions
$J(Q\Lambda)$.

For power counting operators in \SCETb it is useful to have a general formula
that applies both to external operators and soft-collinear Lagrangians, much
like the formula~\cite{bpspc} $\delta = 4 + 4u + \sum_k (k-4) V_k^C + (k-8)
V_k^U$ in \SCETa.  Since the full \SCETb formula does not appear in the
literature I take this opportunity to derive it. Counting the factors
$\eta=\Lambda/Q$ in an arbitrary diagram we find it scales as $\sim \eta^\delta$
where
\begin{eqnarray}
  \delta = \sum_k (V_k^C + V_k^{SC} + V_k^S ) + 4 L^C + 4 L^S + 5 L^{SC} 
    - 4 I^C - 4 I^S \,.
\end{eqnarray}
Here $V_k^{j}$ counts the number of type-$j$ operators of order $\eta^k$ (from
the scaling of fields and derivatives). For instance $(\bar\xi_nW)\Gamma
(S^\dagger h_v)$ counts as $V_{5/2}^{SC}=1$. The $L^j$'s count factors of $\eta$
in loop measures, and the $I^j$'s correct for the $\eta$'s from internal
propagators.  Note that in general momentum conservation requires that some
loops will be neither collinear or soft, but instead have momenta
$(p^+,p^-,p^\perp)\sim (\eta^2,\eta,\eta)$. These momenta are counted by
$L^{SC}$ and occur despite the fact that no degrees of freedom (or individual
propagators) have this scaling.  Using the Euler identity $\sum_k (V_k^C +
V_k^{SC} + V_k^S ) + L^S + L^C + L^{SC} - I^S - I^C =1 $ then leaves
\begin{eqnarray} \label{eq1}
  \delta = 4+ \sum_k (k-4)(V_k^C + V_k^{SC} + V_k^S ) +  L^{SC} \,.
\end{eqnarray}
This formula is still inconvenient since it depends on $L^{SC}$ and not solely
on the vertices. This can be circumvented by using two more Euler identities ala
Ref.~\cite{LMR}.  In any diagram if we erase all collinear lines then we are
left with $N_S$ connected components (or $N_C$ if we instead erase all soft
lines). Thus, $\sum_k (V_k^S + V_k^{SC}) + L^S-I^S =N_S$ and $\sum_k (V_k^C +
V_k^{SC}) + L_C - I_C = N_C$, which taken together with our original Euler
identity give
\begin{eqnarray} \label{eq2}
  L^{SC} = 1-N_c-N_s + \sum_k V_k^{SC} \,.
\end{eqnarray}
Note that diagrams with more than one soft-collinear vertex are required for
$L^{SC}\ge 1$.  Combining Eqs.~(\ref{eq1}) and (\ref{eq2}) gives the final
result
\begin{eqnarray} \label{pcII}
   \delta = 5 -N_c -N_s + \sum_k (k-4) (V_k^S + V_k^C) + \sum_k (k-3) V_k^{SC}.
\end{eqnarray}
The factor of $(k-3)$ for subleading mixed soft-collinear Lagrangians indicates
that we should subtract $3$ from the scaling of the operator in determining the
power to use for these Lagrangians. This $-3$ subtraction agrees with
Ref.~\cite{HN}. With the prefactor in Eq.~(\ref{pcII}) the result also works for
external operators.  For instance for $\bar B^0 \to D^+\pi^-$ the electroweak
Hamiltonian matches onto an $\SCETb$ operator that counts $V_5^{SC}=1$,
$5-N_C-N_S=3$, so the leading order diagrams all have $\delta=5$, as in
~\cite{bpspc}.

SCET has rich applications to phenomenology which I do not have room to discuss.
A few examples of processes which have been considered are summarized in
Table~\ref{tab:pc} along with references to the literature. Also shown are the
degrees of freedom and non-perturbative functions that describe each process at
leading order in the power counting.
\begin{table}[t]
\caption{A few examples of processes that can be described by SCET.\label{tab:pc}}
\vspace{0.4cm}
\begin{center}
\begin{tabular}{|l|l|l|l|l|}
\hline
 Process & Frame & Degrees of Freedom ($p^2$) & Non-Pert.~functions & Refs. 
  \\ \hline
 $\bar B^0\to D^+ \pi^-,\ldots$ & $B$ rest & collinear (${\Lambda^2}$), 
   soft (${\Lambda^2}$) & $\xi(w)$, $\phi_\pi$ & \cite{bps}\\
 $\bar B^0\to D^0 \pi^0,\ldots$ & $B$ rest & collinear (${\Lambda^2}$), 
   soft (${\Lambda^2}$) & $S(k_j^+)$, $\phi_\pi$ & \cite{mps}\\
 &  & collinear ($Q\Lambda$) & & \\
 $B\to X_s^{endpt}\gamma$, & $B$ rest  & collinear (${Q\Lambda}$), 
   usoft (${\Lambda^2}$) & $f(k^+)$ & \cite{bfl,bfps}
   \\
 \ $B\to X_u^{endpt}\ell\nu$ & & &  & \\
 $B\to \pi\ell\nu,\ldots$ & $B$ rest  & collinear (${Q\Lambda}$), 
   soft (${\Lambda^2}$), & $\phi_B(k^+)$, $\phi_\pi(x)$, $\zeta_\pi(E)$ 
   & \cite{bps4,ps1} \\
 & & collinear ($\Lambda^2$) & & \cite{bcdf,bfps,bf}\\
 $B\to \gamma\ell\nu$  & $B$ rest  & collinear (${Q\Lambda}$), 
   usoft (${\Lambda^2}$)  & $\phi_B$ & \cite{bgamenu}
   \\
 $B\to \pi\pi$ & $B$ rest & collinear (${Q\Lambda}$), soft (${\Lambda^2}$), 
    & $\phi_B$, $\phi_\pi$, $\zeta_\pi(E)$ & \cite{bpipi} \\
 & & collinear ($\Lambda^2$) & & \\
 $B\to K^*\gamma$ & $B$ rest &  collinear (${Q\Lambda}$), soft (${\Lambda^2}$), 
  & $\phi_B$, $\phi_K$, $\zeta^{\perp}_{K^*}(E)$ & \cite{bKgamma} \\
 & & collinear ($\Lambda^2$) & & \\
 $e^- p\to e^- X$ & $p$ Breit  & collinear (${\Lambda^2}$)
   & $f_{i/p}(\xi)$, $f_{g/p}(\xi)$ &  \cite{bfprs} \\
 $e^-\gamma\to e^-\pi^0$ &  $\pi$ Breit  & collinear (${\Lambda^2}$), 
   soft (${\Lambda^2}$) & $\phi_\pi$ & \cite{bfprs,ira}
   \\
 $e^+e^-\to j_1+jets$ & $e^+e^-$ CM & collinear ($Q\Lambda$), usoft ($\Lambda^2$)
   & ${\cal A}_1^q$, ${\cal A}_1^{g}$ & \cite{wise}  \\
 $\gamma^* M \to M'$ &  $\pi$ Breit  & collinear (${\Lambda^2}$), 
   soft (${\Lambda^2}$) & $\phi_M$, $\phi_{M'}$ & \cite{bfprs,ira}
   \\
 $e^+e^- \to J/\psi X^{endpt}$ & $e^+e^-$ CM & collinear ($Q\Lambda$), usoft
 ($\Lambda^2$) & $S^{(8,n)}(k^+)$ & \cite{Jpsi} \\
 $\Upsilon \to X^{endpt} \gamma$ & $\Upsilon$ rest & collinear ($Q\Lambda$), usoft
 ($\Lambda^2$) & $S^{(8,n)}(k^+)$ & \cite{Upsilon} \\
 \hline
\end{tabular}
\end{center}
\end{table}

\section*{Acknowledgments}
 This work was supported in part by the U.S.  Department of
Energy (DOE) under the cooperative research agreement DF-FC02-94ER40818.

\end{document}